# How reliable are the HQET-sum rule predictions?

S. Narison

Theoretical Physics Division, CERN
CH - 1211 Geneva 23
and
Laboratoire de Physique Mathématique
Université de Montpellier II
Place Eugène Bataillon
34095 - Montpellier Cedex 05

## Abstract

We test the internal consistencies and the reliability of the existing estimates of the decay constant $f_B$ in the static limit, the meson-quark mass gap $\bar\Lambda$ and the kinetic energy $K$ of a heavy quark obtained from the heavy quark effective theory (HQET)-sum rules. Finite energy local duality sum rules (FESR) have also been used to fix *approximatively* the value of the continuum energy and to study the correlations among these different parameters. Then, we deduce to two-loop accuracy: $\bar\Lambda = (0.65 \pm 0.05)$ GeV, $K = -(0.5 \pm 0.2)$ GeV$^2$, implying the value of the pole mass in HQET: $M_b = (4.61 \pm 0.05)$ GeV. By combining the results from the sum rules in HQET and in the full theory, we obtain $f_B^\infty = (1.98 \pm 0.31)f_\pi$ and the quark mass dependence of the pseudoscalar decay constant: $f_P\sqrt{M_P} = (0.33 \pm 0.06)$ GeV$^{3/2}\alpha_s^{1/\beta_1}\left\{1 - 2\alpha_s/3\pi - 1.1/M_Q + 0.7/M_Q^2\right\}$.

There has been a lot of effort during these last two years for estimating the parameters of the heavy quark effective theory (HQET) [1] using QCD spectral sum rules (QSSR) [2, 3] in the infinite mass limit [4]–[14]. Among other existing estimates, we shall focus our discussion on the determinations of the decay constant $f_B$ in the static limit, the meson-quark mass gap $\bar{\Lambda}$ and the kinetic energy $K$ of a heavy quark. Our aim is to test the reliability of the existing estimates of these quantities and to try to extract the most reliable information from the sum rules analysis. In so doing, we shall first reconsider the HQET sum rules analysis by using an update of the QCD parameters and by testing the stability of the results with the variation of the Laplace sum rule variable $\tau$ and/or of the continuum threshold $E_c$ [1]. If this stability exists, then, we shall compare it with the corresponding *conservative* values of the set $(\tau, E_c)$ obtained from QCD spectral sum rules (QSSR) in the full theory [2]. We shall also use local duality finite energy sum rule (FESR) to fix *approximatively* $E_c$ and study the different correlations among these different parameters.

## 2 Definitions and normalizations

The decay constant of the pseudoscalar meson is defined as:

$$\langle 0|J_5^\mu|B\rangle = \sqrt{2} f_B q^\mu \qquad f_\pi = 93.3 \text{ MeV}, \tag{1}$$

where $J_5^\mu$ is the axial current of the full QCD theory. In HQET, the decay constant reads:

$$\langle 0|\tilde{J}_5^\mu|B\rangle = \frac{i}{\sqrt{2}} \tilde{f}_B v^\mu, \tag{2}$$

to which corresponds the RGI coupling:

$$\hat{f}_B = \tilde{f}_B(\nu) (\alpha_s)^{1/-\beta_1} \left(1 - \delta \frac{\alpha_s(\nu)}{\pi}\right), \tag{3}$$

where $\beta_1 = -1/2(11 - 2n_f/3)$ is the first coefficient of the $\beta$-function and $\delta \simeq -0.23$ for $n_f = 4$ flavours; $v^\mu$ is the heavy quark velocity; $\tilde{J}_5^\mu$ is the axial current in HQET, which is related to the former $J_5^\mu$ of the full theory by:

$$J_5^\mu \simeq \tilde{J}_5^\mu \left(1 - \frac{2}{3} \frac{\alpha_s(\nu)}{\pi}\right), \tag{4}$$

---

[1] The meaning of this stability has been discussed in [9] in quantum mechanics, by taking the Laplace transform of the eigenvalue solution of an harmonic oscillator and by comparing the approximate series with the known complete solution. The optimal result is obtained at the minimum or at the inflexion point of the curves.

[2] These conservative values correspond to the value of $E_c$ comprised between that at which the $\tau$-stability is reached and that where we have $E_c$-stability. These criteria include *all* different alternative criteria used in the literature and gives a more precise meaning of the so-called sum rule window of SVZ.

$$\left(f_B\sqrt{M_B}\right)^{\infty} \simeq \tilde{f}_B \left(\alpha_s(M_b)\right)^{1/\beta_1} \left(1 - \frac{2}{3}\frac{\alpha_s(M_b)}{\pi}\right), \tag{5}$$

The meson-quark mass gap $\bar{\Lambda}$ is defined as [1] [3]:

$$M_B = M_b + \bar{\Lambda} - \frac{1}{2M_b}\left(K + 3\Sigma\right), \tag{6}$$

where:

$$K = \frac{1}{2M_B}\langle B(v)|\mathcal{K}|B(v)\rangle \qquad \text{and} \qquad \Sigma = \frac{1}{6M_B}\langle B(v)|\mathcal{S}|B(v)\rangle \tag{7}$$

corresponds respectively to the matrix elements of the kinetic and chromomagnetic operators:

$$\mathcal{K} \equiv \bar{h}(iD)^2 h \qquad \text{and} \qquad \mathcal{S} \equiv \frac{1}{2}\bar{h}\sigma_{\mu\nu}F^{\mu\nu}h, \tag{8}$$

where $h$ is the heavy quark field and $F^{\mu\nu}$ the electric field tensor.

## 3   HQET sum rules

Let us start our discussion from the relativistic Laplace sum rule (LSR) [2, 3]:

$$\mathcal{L}_R(\tau_r) = \int_0^{t_c} dt \; e^{-t\tau_r} \frac{1}{\pi}\text{Im}\Pi(t), \tag{9}$$

where $t_c$, $\tau_r$ are respectively the QCD continuum threshold and the Laplace sum rule variable, which *a priori* are free external parameters in the sum rule analysis; $\Pi$ is a generic notation of the hadronic Green's function. HQET sum rules correspond to the limit of LSR where the heavy quark mass tends to infinity. In this case, it is more convenient to introduce the non-relativistic variables $\tau$ and $E$, which are related to the former as:

$$t \equiv (E + M_Q)^2, \qquad \tau_e \equiv \tau_r \frac{M_Q}{2}\left(1 + \mathcal{O}\left(\frac{1}{M_Q}\right)\right), \tag{10}$$

or:

$$\omega \equiv 2E, \qquad \tau_\omega \equiv \tau_r M_Q, \tag{11}$$

which remain finite when $M_Q$ tends to infinity. These non-relativistic sum rules have been discussed earlier [9, 10], and have been exploited in the context of HQET in [4]–[8]. They have also been used in full QCD but for increasing the value of $M_Q$ in [11]–[14] in order to reach the HQET regime from below. It has also been shown [12] that the Laplace and moments sum rules are equivalent, when the number of derivatives $n$ of the $n$-th moments and $M_Q$ are large, but their ratio $\tau \equiv n/M_Q$ remains finite [4]. Therefore, without loss of

---

[3] We are aware of the fact that in the lattice calculations, $\bar{\Lambda}$ defined in this way can be affected by renormalons [15].

[4] However, it has to be noticed that in the analysis of the heavy-to-light transitions, the optimal results are obtained at small $n$-values [17], such that, the Moments sum rule does not tend to the Laplace one, when $M_Q$ is large.

rule:

$$\mathcal{L}(\tau_e) = \int_0^{E_c} dE \; e^{-E\tau_e} \frac{1}{\pi} \mathrm{Im}\Pi(E), \tag{12}$$

or equivalently with:

$$\mathcal{L}(\tau_\omega) = \int_0^{\omega_c} d\omega \; e^{-\omega\tau_\omega} \frac{1}{\pi} \mathrm{Im}\Pi(\omega); \tag{13}$$

both sum rules have been used in the literature. The apparent advantage of Eq. (13) is that it optimizes at a value of $\tau_\omega \simeq \tau_e/2$, which is a trivial fact due to their intrisic definitions, but it is *psychologically* more comfortable to reach the stability at a value of $\tau^{-1}$ around 1 but not 2 GeV$^{-1}$, therefore giving a better justification for the convergence of the Operator Product Expansion.

## 4 The mass gap $\bar{\Lambda}$ and the decay constant $\tilde{f}_B$

The sum rule for the decay constant and for the mass gap $\bar{\Lambda}$ is well-known in the literature. It reads [4]–[6], [3, 12]:

$$
\begin{aligned}
2\tilde{f}_B^2 e^{-2\bar{\Lambda}\tau} &\simeq \frac{3}{8\pi^2} \int_0^{\omega_c} d\omega \; \omega^2 \; e^{-\omega\tau} \left\{ 1 + \frac{\alpha_s(\nu)}{\pi} \left( \frac{17}{3} + \frac{4}{9}\pi^2 - 4\ln\omega/\nu \right) \right\} \\
&\quad - \langle \bar{u}u \rangle(\nu) \left\{ 1 + \frac{\alpha_s(\nu)}{\pi} \left( 2 + \frac{\Delta}{16\beta_1} \right) - \frac{M_0^2}{4} \tau^2 \alpha_s^{-2/(3\beta_1)} \right\},
\end{aligned}
\tag{14}
$$

where $\tau \equiv \tau_\omega$; $\Delta = 704/9 - 112\pi^2/27$ [5], but its effect is not significant in the analysis. The subtraction point $\nu$ can be naturally chosen at $\nu = 1/\tau$, as dictated by the RGE which governs the Laplace sum rule [18]. However, in order to study the $\nu$-dependence of the results, we shall take it from 1 GeV to $2/\tau_e$, which covers the choice done in the literature. The mass gap can be deduced from the log-derivative of Eq. (14) with respect to $\tau$. We shall use the values of the QCD scale for 4 flavours to two-loops:

$$\Lambda_4 = (317 \pm 84) \text{ MeV}, \tag{15}$$

corresponding to $\alpha_s(M_Z) \simeq 0.118 \pm 0.006$ [19], and of the QCD non-perturbative parameters [3]:

$$M_0^2 = (0.8 \pm 0.01) \text{ GeV}^2 \quad \langle \bar{u}u \rangle (1 \text{ GeV}) \simeq -(0.223 \text{ GeV})^3 \quad \langle \alpha_s G^2 \rangle = (0.06 \pm 0.03) \text{ GeV}^4. \tag{16}$$

We study the $\tau$-behaviours of $\tilde{f}_B$ and of $\bar{\Lambda}$ in Figs. 1a, c for a fixed value of $E_c$. At the approximation where the correlator is computed, there is a good stability for the values of $\tau$ in the range of 1 to 2 GeV$^{-1}$. The dependence of these optima on $E_c$ are shown in Fig. 1b, d, where $\tilde{f}_B$ has a much better stability than $\bar{\Lambda}$. Taking the *conservative* range of $E_c$ suggested by the analysis of $f_B$ from QSSR in the full theory [3], [11]–[14]:

$$E_c \simeq 1 \sim 1.6 \text{ GeV}, \tag{17}$$

$$\tilde{f}_B \simeq (0.21 \sim 0.28) \text{ GeV}^{3/2} \qquad \bar{\Lambda} \simeq (0.52 \sim 0.70) \text{ GeV}. \qquad (18)$$

The weak dependence of the results on $E_c$ shows the irrelevance of the criticisms of [7] on the choice of $E_c$ done in [11], while the ones of [16] do not concern the working regions of the sum rules. It should be noticed that the effect of the subtraction scale in the range proposed previously is not important. Varying $\alpha_s$ within the error bar given previously, one affects $\bar{\Lambda}$ and $\tilde{f}_B$ respectively by 2 and 7%. The error due to the quark condensate is much smaller. The most significant error not taken into account in the HQET-literature with the exception of [12] is the one due to the unknown $\alpha_s^2$-term. This is important for $\tilde{f}_B$ as the perturbative radiative correction is huge, which is of the order of the lowest-order term in the two-point correlator. In the absence of an evaluation of this term, a reasonable estimate is obtained by assuming that its coefficient grows like a geometric sum or, more conservatively that its effect cannot be as large as the known $\alpha_s$-term. In this way, an error of about 25 % on $\tilde{f}_B$ is introduced [5]. Then:

$$\tilde{f}_B \simeq (0.25 \pm 0.04 \pm 0.06) \text{ GeV}^{3/2}, \qquad (19)$$

where the first error is due to the sum rule procedure and the second one is mainly dominated by the unknown $\alpha_s^2$-term. The values of $\tilde{f}_B$ and of $\bar{\Lambda}$ are in agreement with the previous results in [5, 6], but the error for $\tilde{f}_B$ must have been underestimated there. We show in Fig. 1d the result of [6] (dashed line). The small difference comes from the different value of $\Lambda_4$ used in the analysis. Moreover, these values are also in agreement with the ones from the sum rule in the full theory, obtained by increasing the quark mass-value [12]:

$$\tilde{f}_B \simeq (0.33 \pm 0.06) \text{ GeV}^{3/2}, \qquad \bar{\Lambda} \simeq (0.6 \sim 0.8) \text{ GeV}, \qquad (20)$$

where $\tilde{f}_B$ corresponds to:

$$\left( f_B \sqrt{M_B} \right)^\infty \simeq (0.42 \pm 0.07) \text{ GeV}^{3/2}. \qquad (21)$$

## 5 The kinetic energy $K$

The sum rules for the kinetic energy have been proposed in [6, 8], where the two-point and three-point functions have been used respectively. However, although the two-point function is a convenient quantity to work with, one cannot disentangle there the different contributions, which are of the same order in $1/M_Q$. Therefore, we shall only consider the three-point function analysis from [8]. The sum rule reads:

$$\tilde{f}_B^2 \ K \ e^{-\bar{\Lambda}\tau} \simeq \int_0^{E_c} dE \int_0^{E_c} dE' \ e^{-(E+E')\tau/2} \Biggl\{ \Biggl\{ -\frac{3}{\pi^2} E^4 \delta(E - E') \\ \times \left[ 1 + \frac{\alpha_s(\nu)}{\pi} \left( \frac{41}{9} + \frac{4\pi^2}{9} - 2\ln E\tau \right) \right] -$$

---
[5] In the following this error will be systematically included.

$$-\frac{}{\pi}\frac{}{\pi^2}(E+E')(E^2+E'^2-|E^2-E'^2|)\Bigg\}$$
$$+\frac{4}{3}\frac{\alpha_s(\nu)}{\pi}\langle\bar{u}u\rangle(\nu)\Big[E\delta(E')+E'\delta(E)+(E+E')\delta(E-E')\Big]$$
$$-\frac{1}{4\pi}\langle\alpha_s G^2\rangle\delta(E-E')+\frac{3}{8}\alpha_s^{-2/(3\beta_1)}M_0^2\langle\bar{u}u\rangle(\nu)\delta(E)\delta(E')\Bigg\},$$
(22)

where $\tau \equiv \tau_e$. The kinetic energy $K$ can be obtained by taking the ratio of the sum rules in Eqs. (22) and (14). A similar analysis can be done here, which is summarized in Fig. 2a, b. Again a good stability in $\tau$ is reached for $E_c \geq 0.8$ GeV. This suggests the *phenomenological* but not rigorous lower bound:

$$|K| \geq 0.4 \text{ GeV}^2. \tag{23}$$

However, the dependence on $E_c$ is *violent*. Hopefully, a *slight* inflexion point appears for $E_c \simeq 1 \sim 1.2$ GeV, at which one can extract the optimal result:

$$K \simeq -(0.5 \sim 0.6) \text{ GeV}^2, \tag{24}$$

but our confidence level in this estimate is less than for the previous parameters. A more conservative attitude is to consider the whole range spanned in Eqs. (23) and (24) and to enlarge the error by a factor 2. Thus, one obtains:

$$K \simeq -(0.5 \pm 0.2) \text{ GeV}^2. \tag{25}$$

Though in agreement with that of [8], this result suggests that a lower value of $K$ about -0.3 GeV$^2$ is not absolutely excluded.

## 6  Local duality FESR constraints

Let us now test the consistency of the previous estimates by using local duality FESR. The constraints from the lowest FESR moments can be derived from the LSR by taking the limit $\tau \to 0$. In this way, one can deduce the set of *approximate* equations:

$$\begin{aligned}
\bar{\Lambda} &\approx \frac{3}{4}E_c\left\{1+\frac{\pi^2}{E_c^3}\langle\bar{u}u\rangle\right\} \\
2\tilde{f}_B^2 &\approx \frac{1}{\pi^2}E_c^3\left\{1+\frac{\alpha_s}{\pi}\left(\frac{17}{3}+\frac{4}{9}\pi^2\right)-\frac{\pi^2}{E_c^3}\langle\bar{u}u\rangle\right\} \\
-K &\approx \frac{3}{5}E_c^2\left\{1-\frac{10}{9}\frac{\alpha_s}{\pi}+\frac{\pi^2}{E_c^3}\langle\bar{u}u\rangle\left(1-\frac{5}{8}\frac{M_0^2}{E_c^2}\right)\right\}.
\end{aligned} \tag{26}$$

We solve these constraints by using as input the value of $\bar{\Lambda}$ obtained previously, which is quite stable versus the variation of $E_c$ and which is also less affected by the radiative correction than $\tilde{f}_B$. This gives:

$$E_c \approx 1 \text{ GeV} \tag{27}$$

$$\tilde{f}_B \approx (0.40 \pm 0.07) \text{ GeV}^{3/2}$$
$$K \approx -0.41 \text{ GeV}^2, \tag{28}$$

which agrees within the errors with the previous estimates and indicates the consistency of the whole results from the LSR and FESR.

# 7 Conclusions

Using the LSR and local duality FESR, we have derived the values of the HQET parameters, at which the results are consistent with each other. We conclude that, besides the eventual contribution of renormalons, the most reliable estimate is that of $\bar{\Lambda}$, although it is slightly affected by the variation of $E_c$. Indeed, contrary to $\tilde{f}_B$, its radiative correction is quite small (ratio of sum rules). We can consider as the *best* estimate, the interaction region between the HQET and full QCD sum rules results in Eqs. (18) and (20), which suggests:
$$\bar{\Lambda} \simeq (0.65 \pm 0.05) \text{ GeV}. \tag{29}$$

This value is slightly higher than the existing estimates [6, 5] and the recent average [22], which is based on a more precise value $\bar{\Lambda} \simeq (0.55 \pm 0.05)$ GeV from HQET sum rule than the one in Eq. (18). For the kinetic energy, the result is inaccurate due to the huge dependence of the estimate on the change of the value of $E_c$. Our *conservative* estimate in Eq. (25), which spans the range of values provided by the LSR and FESR methods is [6]:
$$K \simeq -(0.5 \pm 0.2) \text{ GeV}^2. \tag{30}$$

By combining the previous results with the well-known estimate of the chromomagnetic matrix element to leading order in $1/M_b$:
$$\Sigma \simeq \frac{1}{4}(M_{B^*}^2 - M_B^2) = 0.119 \text{ GeV}^2, \tag{31}$$

one can deduce from Eq. (6), the value of the quark mass to two-loop accuracy entering in HQET:
$$M_b \simeq (4.61 \pm 0.05) \text{ GeV}. \tag{32}$$

This value is in agreement (within the errors), with previous analysis from HQET-sum rules[6] and with the most recent determinations of the pole mass to two loops from the relativistic sum rules fit of the $B$ and $B^*$ masses in the full theory [22]:
$$M_b \simeq (4.63 \pm 0.08) \text{ GeV}, \tag{33}$$

and from a non-relativistic sum rule analysis of the bottomium systems:
$$M_b \simeq (4.69 \pm 0.03) \text{ GeV}, \tag{34}$$

---
[6] We are aware of the importance of this quantity in the analysis of the inclusive $B$-meson decays [21].

contributions in $1/M_b$ to the relation given in Eq. (6). It may also indicate that the still quantitatively less controlled renormalon effects to the quark mass defined in HQET[23], is small regardless the definition of $\bar{\Lambda}$ introduced in the analysis. [7]. For the decay constant $\tilde{f}_B$, the radiative correction is huge, and it amounts to about 100% of the lowest-order contribution in the corresponding two-point correlator. Therefore, in the absence of an evaluation of the $\alpha_s^2$-term, one can only suggest the *conservative* range spanned by the results of the two different HQET sum rules. Then, we obtain

$$\tilde{f}_B \simeq (0.34 \pm 0.16) \text{ GeV}^{3/2}, \tag{35}$$

where, as already mentioned, the error is mainly due to the estimate of the unknown $\alpha_s^2$-corrections. This result can be transformed into the value of the decay constant in the so-called static limit, by using the relation in Eq. (5). Therefore, one obtains:

$$f_B^\infty \simeq (2.02 \pm 0.9) f_\pi. \tag{36}$$

This result is in agreement with the value:

$$f_B^\infty \simeq (1.97 \pm 0.33) f_\pi \tag{37}$$

obtained from the sum rule in the full theory [12], which does not however suffer from a large $\alpha_s$-correction and therefore has a much smaller error. One can combine these two results and quote the *final value* of the decay constant in the static limit from the sum rules:

$$f_B^\infty \simeq (1.98 \pm 0.31) f_\pi \quad \text{and} \quad \tilde{f}_B \simeq (0.33 \pm 0.06) \text{ GeV}^{3/2}. \tag{38}$$

This result is in good agreement with the present lattice results $f_B^\infty \simeq (1.74 \pm 0.27) f_\pi$ compiled by [20]. This result can be used for fixing the coefficients $A$ and $B$ of the $1/M_Q$ and $1/M_Q^2$ corrections on $f_P$ from Eq. (5):

$$f_P \sqrt{M_P} \simeq \tilde{f}_B \left(\alpha_s(M_Q)\right)^{1/\beta_1} \left(1 - \frac{2}{3}\frac{\alpha_s(M_Q)}{\pi} - \frac{A}{M_Q} + \frac{B}{M_Q^2}\right), \tag{39}$$

In so doing, we shall use the values [22, 12, 24]:

$$f_D \simeq (1.37 \pm 0.04 \pm 0.06) f_\pi \quad \text{and} \quad f_B \simeq (1.49 \pm 0.06 \pm 0.05) f_\pi. \tag{40}$$

First, we use a linear fit and fix simply $A$ from the ratio $f_B/f_D$. We obtain:

$$A = (0.6 \pm 0.1) \text{ GeV}, \tag{41}$$

which agrees within the errors with the previous findings [11], [5]–[8], [12]. Secondly, we determine $A$ and $B$ by solving numerically these three equations. We obtain:

$$A \approx 1.1 \text{ GeV} \quad \text{and} \quad B \approx 0.7 \text{ GeV}^2, \tag{42}$$

where $A$ and $B$ have large errors. These two terms agree within the errors with the earlier results [11], which suggests that the $1/M_Q^2$ correction is smaller than the $1/M_Q$ one even at the charm quark mass. been

---

[7] An indication of the smallness of this effect can also be inferred from the comparison of the values of the pole mass from relativistic and non-relativistic bottomium sum rules [22]

**Fig. 1** : Analysis of the $\tau$- (**a, c**)) and $E_c$- (**b, d**) behaviours of $\tilde{f}_B$ and $\bar{\Lambda}$ for $\Lambda_4 = 0.317$ GeV. The dashed line is the prediction of [6] using $\Lambda_4 = 0.25$ GeV. **Fig. 2** : The same as Fig. 1 but for $K$. In **a**), we show the effect of the subtraction point.